\documentstyle[12pt]{article}
\def\fnote#1#2{\begingroup\def\thefootnote{#1}\footnote{#2}
\endgroup}

\begin{document}

\hfill {RIMS-1036}

\hfill{UTTG-18-95}

\vspace{24pt}

\begin{center}
{\large{\bf   Are Nonrenormalizable Gauge Theories
Renormalizable?}}

\vspace{36pt}
Joaquim Gomis\fnote{a}{Permanent address: Dept.~d'
Estructura i Constituents de la Mat\`eria, University of
Barcelona; gomis@ecm.ub.es.}\\
Research Institute for Mathematical Sciences\\
Kyoto University, Kyoto 606-01, JAPAN\\

\vspace{12pt}
Steven
Weinberg\fnote{b}{Research
supported in part by the
Robert A. Welch
 Foundation and NSF Grants PHY 9009850 and PHY 9511632.}\\
Theory Group, Department of Physics, University of Texas\\
Austin, TX, 78712, USA\\
weinberg@physics.utexas.edu
\end{center}

\vspace{30pt}

\noindent
{\bf Abstract} --- We raise the issue whether gauge
theories, that are not renormalizable in the usual
power-counting sense, are nevertheless renormalizable in the
modern sense
that all divergences can be cancelled by renormalization of
the infinite number of terms in the bare action.   We find
that a theory is renormalizable in this sense if the {\em a
priori}
constraints that we impose on the
form of the bare action correspond to the cohomology of the
BRST transformations generated by the action.  Recent
cohomology theorems of Barnich, Brandt, and Henneaux are
used to show that conventionally nonrenormalizable theories
of Yang-Mills fields (such as quantum chromodynamics with
heavy quarks integrated out) and/or gravitation are
renormalizable in the modern sense.

%
%
\vfill

\baselineskip=24pt
\pagebreak
\setcounter{page}{1}

\noindent
{\bf 1. Introduction}
\nopagebreak

There are two senses in which we may say that a theory is
perturbatively renormalizable.  The first is that the theory
satisfies the
old Dyson criterion, that the Lagrangian density should
contain only operators of dimensionality four or less.$^1$
This condition is a necessary (though not sufficient)
requirement for infinities to be cancelled with only a
finite number of terms in the Lagrangian.   Even with this
condition violated, it still may be possible that all
divergences are cancelled by renormalization of the terms in
the Lagrangian, but that an infinite number of terms are
needed.  Despite the presence of an infinite number of free
parameters, such theories have a good deal of predictive
power --- specifically, all the predictive power in the
$S$-matrix axioms of unitarity, analyticity, etc., together
with
whatever symmetries are imposed on the theory --- and can be
used to carry out useful perturbative calculations.$^2$

Today it is widely believed that all our present realistic
field theories are actually accompanied by interactions that
violate the Dyson criterion.  The standard model is
presumably what we
get when we integrate out modes of very high energy from
some unknown theory, perhaps a string theory, and like any
other effective field theory its Lagrangian density contains
terms of arbitrary
dimensionality, though the terms in the
Lagrangian density with dimensionality greater than four are
suppressed by negative powers of very large masses.
Likewise for general relativity; there is no reason to
believe that the Einstein-Hilbert action is the whole story,
but all terms in the action with more than two derivatives
are suppressed by negative powers of a very large mass,
perhaps
the Planck mass.  Even if
we were to take seriously the idea that, say, the strong
interactions are described by a fundamental gauge theory
whose Lagrangian contains only terms of dimensionality four
or less, nevertheless in calculations of processes at a few
GeV we would use an effective field theory with heavier
quarks
integrated out, and  such an effective theory necessarily
involves terms in the Lagrangian of unlimited
dimensionality.  Similarly, although modern string theories
have been generally based on two-dimensional field theories
that are
renormalizable in the Dyson sense, there is some interest in
including terms in the action that violate this
condition.$^3$

The second, `modern,' sense in which a theory may be said to
be
renormalizable is that the infinities from loop graphs are
constrained by the symmetries of the bare action in such a
way that there is a counterterm available to absorb every
infinity.  Unlike the Dyson criterion, this condition is
absolutely necessary for a theory to make sense
perturbatively.  It is
automatically satisfied if the only limitations imposed on
the terms in the bare action arise from global, linearly
realized symmetries.  The difficulty in satisfying this
condition appears when we impose nonlinearly realized
symmetries or
gauge symmetries on the
bare action.  Nonlinearly realized symmetries of the bare
action are in general not symmetries of the quantum
effective action, while gauge symmetries must be eliminated
in
quantizing the theory.  A  BRST symmetry$^4$
does survive the gauge fixing, but it is nonlinearly
realized, so that even though the quantum effective action
respects a BRST symmetry, it is not the same as the BRST
symmetry of the bare action.

The question of whether gauge theories are renormalizable in
the modern sense was originally answered only  in the
context of theories that are renormalizable in the Dyson
sense.$^5$   These proofs relied on
a
brute force enumeration of the possible terms in the quantum
effective action of dimensionality four or less, and it was
not
obvious that these proofs of renormalizability could be
extended to Lagrangian densities that contain terms of
unlimited dimensionality.  This is what is meant by the
question asked in the title of this article.\fnote{*}{To
avoid possible confusion, we should distinguish  between our
aims in this paper and earlier efforts$^6$ to make general
relativity and other theories renormalizable in the Dyson
sense by including higher derivative terms (such as terms
bilinear in the curvature) in the unperturbed Lagrangian.
Such efforts lead to problems with unitarity at the energies
at which the renormalized momentum-space integrals begin to
converge.  In contrast, we accept the conventional way of
splitting
the Lagrangian into unperturbed and interaction terms, so
that the unperturbed Lagrangian correctly describes the
particle content of the theory, and no problems with
unitarity arise in perturbation theory.  Our aim here is not
to restore
renormalizability in the Dyson sense, but to learn how to
live without it.}

Section 2 discusses the `structural constraints' that are
imposed on the bare action in specifying a gauge symmetry.
Section 3 outlines our method for addressing the question of
renormalizability by the use of the antibracket
formalism.$^{7,8}$  We find there that renormalizability in
the modern sense is guaranteed if the structural constraints
imposed on the action are chosen in correspondence with the
cohomology of the antibracket
transformation generated by the bare action.    (The
renormalizability of theories with nonlinearly realized
global symmetries can be dealt with by the same formalism,
but with spacetime-independent ghost fields.)  In section 4
we use  recently proved cohomology theorems$^9$  to show
that theories of Yang-Mills fields and/or gravitation are
renormalizable in the modern
sense, even though we allow terms in the Lagrangian of
arbitrary dimensionality.  But we shall see that the
matching of structural constraints with antibracket
cohomologies is only a sufficient, not a necessary,
condition for renormalizablity.  Cohomology theorems give
the {\em candidates} for ultraviolet divergences or
anomalies; a perturbative calculation is needed to see
whether the divergences or anomalies actually occur.  In
fact, in Section 4 we shall encounter terms in the
cohomology of the antibracket operator that do not
correspond to actual infinities.

There are other cohomology
theorems$^{10}$ that can be applied to `first-quantized'
string
theories.  The
question of the
renormalizability of  supergravity and superstring theories
remains open, but can be studied by the methods of
antibracket cohomology.  It would be reassuring to prove
that all these theories are renormalizable in the modern
sense, but even more interesting if some were not, for then
renormalizability could again be used, as we used to think
that the Dyson power-counting condition could be used, as a
criterion for selecting physically acceptable theories.

Our discussion does not pretend to be mathematically
rigorous.  In particular we work with infinite quantities
without explicit consideration of possible regulators, and
simply assume that there is some way of introducing a
regulator that does not produce anomalies that would
invalidate our arguments.  This is no problem in Yang-Mills
theories that are free of anomalies in one-loop order
because of the nature of
the gauge group rather than because of cancellations among
different fermion multiplets.  In such theories the
cohomology theorem of reference 9 shows that the gauge
symmetries are free of anomalies to all orders, without
regard to the dimensionality of the Lagrangian.  Theories
with $U(1)$ factors may present special difficulties.$^{11}$

Before proceeding, we wish to comment on earlier work on the
renormalization of
general gauge theories, most of which were brought to our
attention after the circulation of an earlier version of
this paper.  Dixon$^{12}$ and then Voronov, Tyutin, and
Lavrov$^{13}$
generalized the ideas of
Zinn-Justin$^7$ by introducing a canonical transformation of
fields and antifields as well as an order-by-order
renormalization of coupling constants.  They emphasized
theories that are renormalizable in the Dyson sense,  but
Voronov, Tyutin, and Lavrov briefly considered more
general theories.    More recently,
Anselmi$^{14}$ has further analyzed the issue of
renormalization in gauge theories that are not
renormalizable in the Dyson sense.  He also uses a canonical
transformation as well as coupling constant renormalization
to cancel infinities, and notes the possibility that
cohomological restrictions might force a weakening of what
we here call `structural constraints,' but his motivation is
different from
ours; he expresses the view that theories with infinite
numbers of free parameters are not `predictive,' and
explains that his purpose is to find a framework for
reducing the infinite number of free parameters in such
theories to a finite number.  Also, Harada, Kugo, and
Yamawaki$^{15}$ have recently studied certain aspects of the
renormalization  of a conventionally non-renormalizable
gauge theory (a gauge-invariant formulation of a non-linear
sigma model), using a generalization of the
Zinn-Justin algorithm.
In contrast with these earlier
references, we aim here at showing how to use gauge theories
with infinite numbers of free parameters as realistic field
theories.  Apart from our different motivation, we also give
a more explicit discussion of the necessity of the possible
structural constraints  imposed on the bare action, which
are used here to deal with the obstructions that arise, for
example, for gauge groups with $U(1)$ factors.  Our
demonstration that renormalizability follows from cohomology
is not limited to any specific choice of structural
constraints, but
only assumes that these are chosen in correspondence with
the infinite terms in the BRST-cohomology of the theory,
whatever that might be.  Where some other assumptions make
this impossible, the theory must be regarded as truly
unrenormalizable.

\vspace{18pt}
\noindent
{\bf 2 Structural Constraints}

Our first step is
to consider how to constrain the
bare action to implement local symmetries.
The bare action is taken to be a local
functional\fnote{**}{In a sense the bare action is not
local, because it is the integral of an infinite power
series in the fields and their derivatives, rather than of a
polynomial in fields and field derivatives.  Bare actions of
this sort may be regarded as perturbatively local, in the
sense that, to any given order of perturbation theory
(whether in small couplings or small energies), only a
finite number of terms in the bare action
contribute.}$S_0[\Phi,\Phi^*]$ of a set of fields $\Phi^n$,
including
some set of `classical' (matter and gauge) fields $\phi^r$,
ghosts $\omega^A$, and perhaps ghosts for ghosts, etc., as
well as
`non-minimal' fields (antighosts $\bar{\omega}^A$, auxiliary
fields $h^A$, and perhaps extraghosts),
and of a corresponding set of antifields $\Phi^*_n$, which
have
statistics opposite to $\Phi^n$.  The bare action is assumed
to satisfy the quantum master equation\fnote{\dagger}{In the
original version of this work, we made the stronger
assumption that both terms in Eq.~(1) vanish.  Both Lavrov
and Tyutin$^{13}$ and Anselmi$^{14}$ considered theories
that satisfy  only the quantum master equation (1).}
\begin{equation}
(S_0,S_0)-2i\hbar \tilde{\Delta} S_0=0\;,
\end{equation}
which incorporates all local symmetries as well as the
associated commutation relations, Jacobi identities,
etc.$^8$
Here $(F,G)$ is
the antibracket
\begin{equation}
(F,G)\equiv \frac{\delta F}{\delta_R\Phi^n}\frac{\delta
G}{\delta_L\Phi_n^*}-\frac{\delta
F}{\delta_R\Phi_n^*}\frac{\delta G}{\delta_L\Phi^n}\;,
\end{equation}
with $L$ and $R$ denoting differentiation from the left and
right, respectively, and $\tilde{\Delta} S_0$ is the
differential operator
\begin{equation}
\tilde{\Delta} \equiv \frac{\delta^2
S_0}{\delta_L\Phi{}^n\,\delta_R\Phi_n{}^*}\;.
\end{equation}
(This is usually called $\Delta$; the tilde is added to
distinguish this from a symbol $\Delta$ introduced later.)
We further suppose that various global,
linearly realized symmetries are imposed, including Lorentz
invariance
and ghost number
conservation.  From now on it should be understood that we
also impose the usual conditions on the antibrackets of the
action with the non-minimal fields $\bar{\omega}^A$ and
$h^A$ and their antifields.

If these were the only
constraints imposed on the action then the theory would
automatically be renormalizable in the modern sense, because
as we shall see in the next section the infinite part
of the quantum effective action in any order would satisfy
the same constraints as the allowed changes in the
counterterms in the bare action.
But not all theories are renormalizable in this sense.  One
very familiar example of a theory that is not
renormalizable in the modern sense is one in which we
arbitrarily set some parameter (such as the
$(\phi^\dagger\phi)^2$ coupling in the electrodynamics of a
charged scalar $\phi$) equal to zero or any finite value.
We are concerned here rather with what we shall
call `structural constraints' --- the constraints that tell
us what gauge symmetries are respected by the theory.

The structural constraints can be of various types:

\vspace{6pt}
\noindent
(a) The usual structural
constraints require the bare action $S_0$ to consist of a
term $I[\phi]$ that depends only on the `classical' (gauge
and
matter) fields and is invariant under some prescribed set of
local symmetry transformations, plus appropriate terms
depending also on a
limited number of antifield field factors, whose number and
structure are
constrained by  the master equation.    For instance, for a
theory with a closed irreducible gauge algebra like
Yang-Mills theory or general relativity the action would be
linear in antifields with one ghost $\omega^A$ and antighost
$\bar{\omega}^A$ for each
gauge symmetry:
\begin{equation}
S_0[\Phi,\Phi^*]=I[\phi]
+\omega^A\,C^r_A[\phi]\,\phi^*_r+\mbox{\small$\frac{1}{2}$}
\omega^A\omega^B\, C^C{}_{AB}[\phi]\,\omega^*_C-
\bar{\omega}^*_A\,h^A\;,
\end{equation}
where $I[\phi]$ is invariant under the infinitesimal
transformation $\phi^r\rightarrow\phi^r+\epsilon^A
C_A^r[\phi]$, and $C^C{}_{AB}[\phi]$ is the structure
constant
for these transformations.  (We are using a `De Witt
notation,' in which indices like $A$ and $r$ include a
spacetime coordinate which is integrated in sums over these
indices.)  For supergravity without auxiliary fields the
action
would be
quadratic in antifields.

\vspace{6pt}
\noindent
b) Instead of imposing a fixed gauge symmetry on a theory,
we can instead impose a symmetry with a fixed number of
generators and fixed commutation relations, but with the
effect of the symmetry transformations on the classical
fields left arbitrary.  For instance, in the case of an
irreducible closed gauge symmetry the action would take the
form (4),  but with the
transformation functions $C^r_A[\phi]$ otherwise
arbitrary.\fnote{\dagger\dagger}{For instance, instead of
the usual
isospin matrices $t_i$ representing the algebra of $SU(2)$
we can take the generators of the $SU(2)$ gauge
transformations to be linear combinations $O_{ij}t_j$.  As
long as the matrix $O_{ij}$ is
real, orthogonal, and  unimodular, this will not change the
$SU(2)$ structure constants.  In this case, the change in
the gauge transformations is the same as would be produced
by a redefinition of the gauge fields.  The cohomology
theorem$^9$ used in Section 4 shows that in
all semisimple Yang-Mills theories and gravitational
theories any infinitesimal change
in the transformation functions $C^r_A[\phi]$ is the same as
would be produced by a redefinition of fields and antifields
together with a corresponding change in $I[\phi]$, but this
is not the
case in general.  For instance, changing the {\em ratios} of
the coupling
constants of various particles to a $U(1)$ gauge field would
change the $U(1)$ transformation rules in a way that could
not be absorbed into a renormalization of the gauge field,
while of course
leaving the structure constants zero.}  This case provides
an illustration of the fact that when we make a change
$\Delta S_0$ in the bare action, the structural
constraints apply to $S_0+\Delta S_0$ rather than to
$\Delta S_0$ itself.  In particular, $\Delta I[\phi]$ is not
necessarily invariant under the original gauge
transformation $\phi^r\rightarrow\phi^r+\epsilon^A
C_A^r[\phi]$, but $I[\phi]+\Delta I[\phi]$ is always
required to be invariant under the transformation
$\phi^r\rightarrow\phi^r+\epsilon^A\,(C^r_A[\phi]+\Delta
C^r_A[\phi])$.

\vspace{6pt}
\noindent
c) We might weaken the structural constraints further,
assuming only that the bare action is a polynomial of a
given order in the antifields.  For instance, if we required
that the action is linear in antifields and involves only
the fields $\phi^r$, $\omega^A$, $\bar{\omega}^A$, and $h^A$
and their antifields, then it would have to take the general
form (4), but with unspecified coefficients $C_r^A[\phi]$
and $C^C{}_{AB}[\phi]$.  In this case the master equation
would require that the action $I[\phi]$ is invariant under
the transformation
$\phi^r\rightarrow\phi^r+\epsilon^AC^r_A[\phi]$ which form
a closed irreducible algebra with structure constants
$C^C{}_{AB}[\phi]$, but we would not be specifying in
advance what this gauge symmetry algebra is or how it is
represented on the matter fields, except in so far as we
specify the transformation of $C_r^A[\phi]$ and
$C^C{}_{AB}[\phi]$ under global linear symmetries.

One convenient aspect of structural constraints of types (a)
and (b) is that we can reverse the connection between the
master equation and the gauge symmetry: an action of the
form (4) will automatically satisfy the quantum master
equation
as long as (1) $I[\phi]$ is invariant under the
transformations
$\phi^r\rightarrow\phi^r+\epsilon^A C_A^r[\phi]$ with
structure constants $C^C{}_{AB}[\phi]$, and (2) a
gauge-invariant regulator is used to define integrals over
fields, so that  $\tilde{\Delta}S_0=0$.  The same is true
when we consider the deformed action $I[\phi]+\Delta
I[\phi]$ and require invariance under the deformed gauge
transformations $\phi^r\rightarrow\phi^r+\epsilon^A
(C_A^r[\phi]+\Delta C_A^r[\phi])$.  This is not true of
structural constraints of type (c); merely assuming that the
action is of some definite order in antifields does not lead
to  the master equation.  We will not need to assume here
that the structural constraints imply the master equation.
We will however assume that (as is true of all the
constraints
discussed above) that the structural constraints are chosen
to be
{\em linear} conditions on possible changes in the action;
if $S_0+A$ and $S_0+B$ both satisfy the structural
constraints, then so does $S_0+\alpha A+\beta B$ for
arbitrary
constants $\alpha$ and $\beta$.  Until Section 4 we will not
be otherwise specific about the structural constraints to be
adopted.

It is these structural constraints
that create a potential problem for renormalizability, for
in general they will not be respected by ultraviolet
divergent terms in the quantum effective action.  The
quantum effective action will not even always satisfy
restrictions on the number of antifield factors,
so that, for example, a bare action with a closed gauge
algebra may yield a quantum effective action with an open
gauge algebra.$^{13}$  Structural constraints arise from our
fundamental assumptions about the sort of theory we wish to
study, but to be physically sensible they must not constrain
a theory so severely that they prevent the cancellation of
ultraviolet divergences.  Our problem is to decide what
structural constraints satisfy this condition.  As we shall
see in the next section, this is a matter of matching the
cohomology of the antibracket operation generated by the
bare action.  Structural constraints of type (a) turn out to
be adequate to deal with general relativity and semisimple
gauge theories.  We would need structural constraints of
type (b) to deal with the candidate divergences that arise
when the gauge group has $U(1)$, but as we shall see these
candidate divergences do not correspond to actual
infinities.  On the other hand, first-quantized string
theories
require structural constraints weaker than those of
type (a).
In considering structural constraints other than those of
type (a) and (b), it is intriguing that here we confront the
possibility that gauge symmetries may be less
fundamental than the antibracket formalism from which they
can be derived.

\vspace{18pt}

\noindent
{\bf 3. Renormalization in General Gauge Theories}
\nopagebreak

We begin with an outline of the antibracket approach to the
renormalization of theories with local symmetries, presented
here in a
way that is independent of the specific structural
constraints imposed on the theory.

\vspace{12pt}
\noindent

\vspace{12pt}
\noindent
A) In analogy with the renormalization of fields in
conventionally renormalizable theories like quantum
electrodynamics, in order for infinities to cancel here we
need
to perform a general canonical transformation
$\Phi\rightarrow \Phi'(\Phi,\Phi^*)$, $\Phi^*\rightarrow
\Phi'{}^*(\Phi,\Phi^*)$ of fields and antifields.  By
an canonical transformation is meant any transformation
that preserves the antibracket structure
\begin{equation}
(\Phi'{}^n,\Phi'_m{}^*)=\delta^n_m\;,\qquad
(\Phi'{}^n,\Phi'{}^m)=(\Phi'_n{}^*,\Phi'_m{}^*)=0\;,
\end{equation}
which insures that antibrackets of general functionals can
be calculated in terms of $\Phi'{}^n$ and $\Phi'_n{}^*$, in
the same way as in terms of $\Phi^n$ and $\Phi^*_n$.
The action $S_0[\Phi,\Phi^*]$ if expressed in terms of the
transformed fields becomes a different functional
$S_0'[\Phi',\Phi'{}^*]\equiv S_0[\Phi,\Phi^*]$, given by
$S'_0[\Phi',\Phi'{}^*]=S_0[\Phi',\Phi'{}^*;1]$, where
$S_0[\Phi,\Phi^*;t]$  is defined by
the differential equation
\begin{equation}
\frac{d}{dt}S_0[\Phi,\Phi^*;t]=\Big(F[\Phi,\Phi^*;t],\,S_0[
\Phi,\,\Phi^*;t]\Big)
\end{equation}
with initial condition
\begin{equation}
S_0[\Phi,\Phi^*;0]=S_0[\Phi,\Phi^*]\;,
\end{equation}
where $F[\Phi,\Phi^*;t]$ is an arbitrary fermionic
functional of ghost number $-1$.
Since
the generator $F$ of the canonical transformation contains
terms
of arbitrary dimensionality, the
bare action $S_0'[\Phi',\Phi'{}^*]$
will  not generally have any simple dependence on the
transformed antifields $\Phi'{}^*$.

\vspace{12pt}
\noindent
B) As a basis for perturbation theory, we must separate out
a finite `renormalized' zeroth-order action $S$ from the
transformed bare
action $S'_0$, with the remainder regarded as a sum of
corrections proportional to powers of
a `loop-counting' parameter $\hbar$, with divergent
coefficients.
The correction term $\Delta S'=S_0'-S$ receives
contributions both
from the counterterm $\Delta S\equiv S_0-S$ in the original
bare action, and also from the
field-antifield-renormalization canonical transformation in
step A.   To be specific, suppose we write the original bare
action as a power series in $\hbar$:
\begin{equation}
S_0=S+\hbar \Delta_1+\mbox{\small$\frac{1}{2}$}\hbar^2
\Delta_2+\cdots\quad.
\end{equation}
The generator $F(t)$ of the canonical transformation
(4) may similarly be written as a power series
\begin{equation}
F(t)=\hbar t F_1+\mbox{\small$\frac{1}{2}$}\hbar^2
t^2F_2+\cdots\;.
\end{equation}
Eqs. (6) and (7) then give the transformed bare action as
\begin{equation}
S_0'=S+\hbar\Big[\Delta_1+(F_1,S)\Big]+\mbox{\small$\frac{1}
{2}$}\hbar^2\Big[\Delta_2+2(F_1,\Delta_1)+(F_2,S)+(F_1,(F_1,
S))\Big]+\cdots\;.
\end{equation}
The renormalized action $S$ is taken to have the same form
as the
original
bare action $S_0$, satisfying the same structural
constraints
(including  the same limitations on its dependence on
antifields), only with finite instead of infinite
coefficients.  Also, since it can be regarded as the limit
of $S_0$ for $\hbar=0$,  it satisfies the classical master
equation
 \begin{equation}
(S,S)=0\;,
\end{equation}
with the antibracket calculated in terms of either the
original or the canonically transformed fields and
antifields.

\vspace{12pt}
\noindent
C) To carry out quantum mechanical calculations of
expectation values, Greens functions, etc., it is necessary
to fix a gauge by taking the antifields as functions of the
fields.  This is usually done by taking the antifields in
the form
\begin{equation}
\Phi^*_n=\frac{\delta \Psi(\Phi)}{\delta\Phi^n}+K_n
\end{equation}
where $\Psi$ is a local fermionic functional of $\Phi$, and
$K_n$ is an external field, held constant in the path
integral.  It is important to recognize that the same
relation then
applies to the transformed antifields
\begin{equation}
\Phi'_n{}^*=\frac{\delta\Psi'(\Phi',K)}{\delta
\Phi'{}^n}+K_n\;,
\end{equation}
but with  a different (and $K$-dependent) gauge-fixing
fermionic functional
$\Psi'$.  We do not know whether a proof of this result has
been published, so a proof is given in an
appendix to this paper.  An observable $O$ will be
unaffected by small changes in $\Psi$, provided it is gauge
invariant, in the sense that $(O,S)-
i\hbar\tilde{\Delta}O=0$.$^{16}$

\vspace{12pt}
\noindent
D) Following the same reasoning as used originally by
Zinn-Justin,$^7$ the quantum effective action
$\Gamma(\Phi,K)$
satisfies the master equation
\begin{equation}
(\Gamma,\Gamma)=0\;,
\end{equation}
 with antibrackets calculated using $K_n$ in place of the
antifield of $\Phi'{}^n$.  But the variables $\Phi'{}^n$ and
$\Phi'_n{}^*$ are related to $\Phi'{}^n$ and $K_n$ by a
canonical transformation, so we can just as well regard
$\Gamma$ as a functional of $\Phi'{}^n$ and $\Phi'_n{}^*$,
satisfying a master equation (14) with the antibracket
calculated in terms of these variables.

In lowest order, $\Gamma$ is the
same as $S$, and is therefore finite.  Suppose that through
cancellations of infinities between loop diagrams and the
counterterm $S_0'-S$, all infinities in $\Gamma$ cancel
up
to some given order $N-1$ in coupling parameters.  Then in
order $N$, the infinite part of the master equation
constrains the infinite part $\Gamma_{N,\infty}$ of the
$N$-th order term in  $\Gamma$ by
\begin{equation}
(S,\Gamma_{N,\infty})=0\;.
\end{equation}
Because $(S,S)=0$, the mapping $X\mapsto (S,X)$
is nilpotent, so that the nature of the solutions of
Eq.~(15) can be determined with the help of appropriate
cohomology theorems.

\vspace{12pt}
\noindent
E)  We shall now suppose that for some given choice of the
structural
constraints discussed in Section 2, we can prove a
cohomology theorem, that any local functional $X$ which is
$S$-closed (in the sense that $(S,X)=0$), and is invariant
under the same linearly realized global symmetries
(including ghost number conservation and Lorentz invariance)
as $S$, may be expressed as
\begin{equation}
X=G+(S,H)
\end{equation}
where $G$ is  a
local functional  for which $S+G$ satisfies the same
structural
constraints as $S$, and $H$ is a local fermionic functional,
with both $G$ and $H$ satisfying the same linearly realized
global symmetries as $S$.   Eq.~(15)
tells us that $\Gamma_{N,\infty}$
is $S$-closed, and it automatically is invariant under the
same linearly realized global symmetries as $S$, so it
satisfies the conditions of this theorem.  The cohomology
theorem will be applied below not to $\Gamma_{N,\infty}$
itself, but to a term in $\Gamma_{N,\infty}$ that also
satisfies these conditions.

Eq.~(10) shows
that in $N$-th order $S_0'$ will
contain  terms  $(F_N,S)$ and $\Delta_N$, which make
additive contributions to $\Gamma_{N,\infty}$, and which do
not depend on the terms in $F$ and $S_0$ that appear in
$\Gamma_M$ for $M<N$.  We must now inquire whether
$\Delta_N$ and $F_N$ can be chosen to cancel the infinities
in $\Gamma_N$.

Because the structural constraints are supposed to be
satisfied by $S_0$ for all $\hbar$, and are assumed to be
linear, they are also satisfied by $S+\Delta_N$.  Now,
apart from these constraints, and
invariance under linearly realized global symmetries, the
only limitation on our freedom to choose the
$N$- \nolinebreak th order
counterterm $\Delta_N$ in the
original bare action is that it should not invalidate the
master equation.  For the structural constraints of type (a)
and (b) discussed in Section 2, this is not much of a
limitation, since the quantum master equation (1)
automatically follows
from these structural constraints, provided we use a gauge-
invariant regulator.  But for future use we
also wish to consider the more general
case, where the master equation must be imposed on $S_0$
independently of the structural constraints.  Since $S_0$ is
supposed to satisfy the master equation for all values of
the loop-counting parameter $\hbar$, the counterterms
$\Delta_N$ are required to satisfy a sequence of equations
\begin{equation}
(S,\Delta_N)=-\mbox{\small$\frac{1}{2}$}\sum_{M=1}^{N-
1}(\Delta_M,\Delta_{N-M})+2i\tilde{\Delta}\Delta_{N-1}\;.
\end{equation}
These conditions on $\Delta_N$ are
{\em not} the same as the
condition $(S,\Gamma_{\infty,N})=0$ on the infinite part of
$\Gamma_N$.

This is no problem.  Suppose we find a
solution of the equations (17) up to order
$N$,\fnote{\ddagger}{The reader may be bothered by the
question of how we know that these
equations can be solved.  It is true that if these equations
are satisfied up to order $N-1$, then the right-hand-side
$R_N$ of the equation for $\Delta_N$ does satisfy the
condition $(S,R_N)=0$, but we cannot find solutions of the
equation $(S,\Delta_N)=R_N$ for arbitrary $R_N$ satisfying
$(S,R_N)=0$ unless the cohomology (known as $H^1(S|d)$,
where $d$ denotes the exterior derivative)  of the
antibracket operation $X\mapsto (S,X)$ on the local
functionals $X$ of ghost number $+1$ is trivial, which is
not generally the case.  (The condition $H^1(S|d)=0$ would
also rule out anomalies,
but it is not a {\em necessary} condition for the theory  to
be
anomaly free.  Even for $H^1(S|d)\neq 0$, anomalies can
cancel among different fermion multiplets, as is the case in
the
standard electroweak theory.)  Fortunately, we are not
trying to solve the equations $(S,\Delta_N)=R_N$ for
arbitrary $R_N$ satisfying $(S,R_N)=0$, but only for the
particular functionals that appear on the right-hand-side of
equations (17).  The existence of such solutions is
guaranteed by the assumption that the structural constraints
allow the master equation to be solved for all values of
$\hbar$.}  which
satisfies the structural constraints.  We may write the
$N$-th order term in the general solution  as
\begin{equation}
\Delta_N=\Delta_N^0+\Delta'_N
\end{equation}
where $\Delta_N^0$ is any particular solution satisfying
Eq.~(17) (and such that $S+\Delta_N^0$ satisfies the
structural constraints), and $\Delta'_N$ is  subject only to
the
conditions that $S+\Delta'_N$ must satisfy the structural
constraints and any linearly realized global symmetries, and
\begin{equation}
(S,\Delta_N')=0\;.
\end{equation}
We may write the infinite $N$-th order terms in $\Gamma$ as
\begin{equation}
\Gamma_{N,\infty}=\Delta'_{N,\infty} -
(S,F_{N,\infty})+X_{N,\infty}
\end{equation}
where $X_N$ consists of terms from loop graphs, as well as
from the term $\Delta_N^0$ and various terms in $\Gamma$
that involve $\Delta_M$ and $F_M$ for $M<N$.  For instance,
for $N=2$ Eq.~(10) gives
\begin{eqnarray*}
X_2&=&\Delta_2^0+2(F_1,\Delta_1)+(F_1,(F_1,S))+{\rm
two\;loop\; terms\;involving\;only}\;S\\&&\qquad+\,{\rm
one\;loop\;terms\;involving}\;S,\;\Delta_1\;{\rm
and}\;F_1\;.
\end{eqnarray*}
For our purposes the only thing we need to know about $X_N$
is that it does not involve $\Delta'_N$ or $F_N$, and that
it is invariant under any linearly realized global
symmetries of $S$.  It follows from Eqs.~(15), (19), and
(20)
that
\begin{equation}
(S,X_{N,\infty})=0\;.
\end{equation}
Hence the hypothesized cohomology theorem would allow  us to
write $X_N$ in the form (16):
\begin{equation}
X_{N,\infty}=G_N+(S,H_N)\;,
\end{equation}
where $G_N$ is  a
local functional  for which $S+G_N$ satisfies the same
structural
constraints as $S$, and $H_N$ is a local fermionic
functional, with both $G_N$ and $H_N$ invariant under the
same linearly realized global symmetries as $S$.
Since $\Delta'_N$ and $F_N$ are local functionals that can
be varied independently of $X_N$, subject only to the
conditions that they are invariant under linearly realized
global symmetries, that $S+\Delta'_N$ satisfies the same
structure constraints,  and that $(S,\Delta'_N)=0$,
they can be chosen so that
\begin{equation}
\Delta'_{N,\infty}=-G_N\;,\qquad\qquad F_{N,\infty}=H_N\;.
\end{equation}
According to Eq.~(20), this eliminates the infinities in the
quantum effective action to order $N$.
 Continuing this process allows a step-by-step construction
of a counterterm $\Delta S$ and canonical transformation
generator $F$ that render the quantum effective action
finite to all
orders.

\vspace{18pt}
\noindent
{\bf 4. Cohomology Theorems}
\nopagebreak

The previous section shows how to use cohomology theorems to
prove the renormalizability of various `nonrenormalizable'
gauge theories.  As an example of such a cohomology theorem,
we note that Barnich, Brandt,  and Henneaux$^9$ have
recently shown that if $S$
is the
action of a {\em semisimple} Yang-Mills theory, or of
gravitation, or both together, which of
course
has ghost number zero and is linear in antifields, then the
most general local functional $X$ of ghost number zero that
satisfies the condition $(S,X)=0$ may be written as a local
gauge-invariant functional $G[\phi]$ of the `classical'
(gauge and matter)
fields alone, so that in our language $S+G[\phi]$ satisfies
the structural
constraints, plus a term of the form
$(S,H)$.
Then by the reasoning of the previous section, we may
eliminate all infinities in the quantum effective action by
adjusting
the counterterms in $S_0-S$ to cancel $G[\phi]$, and
performing a suitable
canonical transformation on the fields and antifields to
cancel $(S,H)$.

Gauge theories with $U(1)$ factors require special
consideration.    Reference 9 shows that in this case the
most general
local functional $X$ of ghost number zero that
satisfies the condition $(S,X)=0$ may be written as a local
gauge-invariant functional $G[\phi]$ of the `classical'
fields alone, plus a term of the form
$(S,H)$, plus a term of the form\fnote{\ddagger\ddagger}{
There are additional complications$^{9}$ in theories with
 certain exotic couplings between matter and
gauge fields.  We will not go into this here, because such
theories do not seem to be of physical interest.}
\begin{equation}
\int A^\mu(x) j_\mu(x)\,d^4x+{\rm terms\; linear\;
in}\;\phi^*_r\;,
\end{equation}
where $j^\mu(x)$ is the gauge-invariant current associated
with any
symmetry of the action, and $A^\mu(x)$ is the $U(1)$ gauge
field (supposing for simplicity that there is only one.)  If
$j_\mu(x)$ is the same current to which $A^\mu(x)$ is
coupled in the bare action, then a term like (24)  can be
compensated by  a renormalization of the field $A^\mu(x)$
and a corresponding renormalization of the antifield
$A_\mu^*(x)$,
which is one example of the canonical transformations
discussed in Step A of the previous section.

On the other hand, if the action respects a global symmetry
in addition to the
$U(1)$ gauge symmetry, then $j^\mu(x)$ can be the current
associated with that global symmetry, and in this case the
cohomology includes terms whose antifield-independent part
is only gauge-invariant `on-shell,' that is, when the field
equations are satisfied.  Thus if infinite terms of the form
(24)
actually appeared in the quantum effective action, with
$j^\mu(x)$ a conserved current other than that to which
$A_\mu(x)$ was originally coupled, then the structural
constraint we used for semisimple gauge theories,
that the bare
action has the form (4) with $I[\phi]$  off-shell invariant
under a prescribed  transformation $\delta\phi^r\rightarrow
\phi^r+\epsilon^A C^r_A[\phi]$, would not lead to a
renormalizable theory.  In this case we
would have to use
the weaker structural constraint of type (b) discussed in
Section 2, that
the action is of the
form (4),
with the transformation functions $C_r^A[\phi]$ specified
only as to their number and structure constants (in this
case zero).
The counterterms in the bare action would then only
be constrained by the condition that they are linear in
antifields, do not invalidate the master equation,
and do not change the structure constants, which in this
case are zero.\fnote{\natural}{As already noted in
Section 2,
the
antifield-independent term $I[\phi]+\Delta I[\phi]$ is not
required by these structural constraints and the master
equation to be invariant
under the original gauge transformations $\phi^r\rightarrow
\phi^r+\epsilon^A C^r_A$, but only under the modified gauge
transformations $\phi^r\rightarrow \phi^r+\epsilon^A
\left(C^r_A+\Delta C^r_A\right)$, so that
$$
\Big(\delta \Delta I[\phi]/\delta \phi^r\Big)  C^r_A=
-\Big(\delta \left(I[\phi]+\Delta I[\phi]\right)/\delta
\phi^r\Big)\,
\Delta C^r_A
$$
which only requires that $\Delta I[\phi]$ should be
invariant under
the original gauge transformation $\phi^r\rightarrow
\phi^r+\epsilon^A C^r_A$ when the field equations are
satisfied.}  Thus such
counterterms could be used to cancel  infinite terms in the
quantum effective action of the form (24).

It does not seem that infinities of the form (24),
with $j^\mu(x)$ a conserved current other than that to which
$A_\mu(x)$ was originally coupled,  actually appear in the
quantum effective action.   We have not checked this by
direct calculation, but such infinite terms would
represent a change in the mixture of fermion currents to
which long-wave photons couple, and this is prohibited by
the Ward soft-photon theorem.  It is not necessary for us to
settle this question, because we have shown that any
infinities of form (24) are cancelled by renormalization of
the parameters in the $U(1)$ gauge transformation, but this
seems to be a case where the candidate divergences presented
by cohomology theorems are not actually divergent.

An even clearer case of this sort is presented by theories
containing  a set of free $U(1)$ gauge fields
$A^b_\mu(x)$.\fnote{\natural\natural}{We are grateful to F.
Brandt
for suggesting this to us.}
The cohomology of the antibracket operator also includes the
terms
\begin{equation}
f_{abc}\int d^x\;\Big(F^{\nu\mu a } A^{b\mu}
A^{c\nu}+2A^{*\mu }_a
A^b_\mu\omega^c+\omega^*_a\omega^b\omega^c\Big)\;.
\end{equation}
where $f_{abc}$ are totally antisymmetric constants.  If
these corresponded to
actual divergences we would have to weaken the structural
constraints so that not even the structure constants were
prescribed in advance, leaving open the possibility that the
fields $A^b_\mu(x)$ transform under a non-Abelian gauge
group.  But here it is quite clear that the terms in
Eq.~(25) are not produced by radiative corrections; no
radiative corrections can give interactions to a field that
does not interact to begin with.

A recent cohomology theorem of
Brandt, Troost, and Van Proeyen$^{10}$ shows that it is also
necessary to weaken the structural constraints in dealing
with first-quantized string theories --- that is, with
gravitation coupled to scalar matter in two dimensions.  If
the Liouville field is explicitly introduced the analysis of
ref. 17 shows that the cohomology of $S$ contains terms
corresponding to a change in the action of its local
symmetries, though not of their algebra, so here one should
impose a structural constraint of type
(b).  Analogous comments apply to the
spinning string.$^{18}$

The
possibility of weakening the structural
constraints may become useful in applications to other
theories.
It is important to find out whether supergravity and general
superstring theories are renormalizable in the modern sense,
and for
this
purpose we need to know the cohomology generated by the bare
action of these theories.

\vspace{12pt}
\noindent
{\bf Acknowledgments}~~~ We are grateful for helpful
conversations with C. Becchi, F. Brandt, D. Buchholz, M.
Henneaux, and J. Pons.

\vspace{18pt}
\noindent
{\bf Appendix}
\nopagebreak

\vspace{12pt}

We wish to prove that if
\begin{equation}
\Phi^*_n=\delta \Psi(\Phi)/\delta\Phi^n+K_n\;,
\end{equation}
 then canonically transformed variables $\Phi'{}^n$ and
$\Phi'_n{}^*$ satisfy a relation of the same form
\begin{equation}
\Phi'_n{}^*=\delta \Psi'(\Phi',K)/\delta \Phi'{}^n+K_n\;,
\end{equation}
though generally with a different (and $K$-dependent)
fermionic functional
$\Psi'\neq
\Psi$.
It is only necessary to show that this is true for
infinitesimal canonical transformations, which are of the
form
\begin{equation}
\Phi'{}^n=\Phi{}^n+(F,\Phi^n)=\Phi{}^n-(\delta F/\delta
\Phi^*_n)_{\Phi^*=\delta\Psi/\delta\Phi+K}\;,
\end{equation}
\begin{equation}
\Phi'_n{}^*=\Phi_n{}^*+(F,\Phi^*_n)=\Phi_n{}^*+(\delta
F/\delta \Phi^n)_{\Phi^*=\delta\Psi/\delta\Phi+K}\;,
\end{equation}
where $F[\Phi,\Phi^*]$ is an infinitesimal fermionic
functional.
Continuity then implies that the same will be true for
finite canonical transformations, in at least a finite
region around the unit transformation.

To prove Eq.~(26), we note that Eqs.~(25) and (28) yield
\begin{equation}
\Phi'_n{}^*=\delta\Psi/\delta \Phi^n+(\delta
F/\delta \Phi^n)_{\Phi^*=\delta\Psi/\delta\Phi+K}+K_n\;.
\end{equation}
The derivative of $\Psi$ with respect to $\Phi$ may be
expressed in terms of its derivative with respect to
$\Phi'$, using Eq.~(27) to write
\begin{equation}
\frac{\delta_L\Phi'{}^n}{\delta\Phi^m}=\delta^n_m-
\frac{\delta_L}{\delta \Phi^m}\left(\frac{\delta F}{\delta
\Phi^*_n} \right)_{\Phi^*=\delta\Psi/\delta\Phi+K}\;.
\end{equation}
Using this in Eq.~(29) and keeping only terms of first
order in $F$ gives
\begin{eqnarray}
\Phi_n'{}^*&=&\frac{\delta\Psi}{\delta \Phi'{}^n}
-\frac{\delta\Psi}{\delta
\Phi^m}\frac{\delta_L}{\delta\Phi^n}\left(\frac{\delta F}{
\delta \Phi_m^*}\right)_{\Phi^*=\delta\Psi/\delta\Phi+K}
+\frac{\delta
F_{\Phi^*=\delta\Psi/\delta\Phi+K}}{\delta\Phi^n}
\nonumber\\&&-
\left(\frac{\delta_L}{\delta\Phi^n}\frac{\delta\Psi}{\delta
\Phi^m}\right)\left(\frac{\delta F}{\delta
\Phi_m^*}\right)_{\Phi^*=\delta\Psi/\delta\Phi+K}+K_n\;.
\end{eqnarray}
To first order in $F$ this has the same form as the desired
result (26), with
\begin{equation}
\Psi'=\Psi-\frac{\delta
\Psi}{\delta\Phi^m}\left(\frac{\delta F}{\delta
\Phi_m^*}\right)_{\Phi^*=\delta\Psi/\delta\Phi+K}+
(F)_{\Phi^*=\delta\Psi/\delta\Phi+K}\;.
\end{equation}

\vspace{36pt}

\noindent
{\bf References}
\nopagebreak
\begin{enumerate}
\item F.J. Dyson, {\it Phys. Rev.} {\bf 75} (1949), 486,
1736.

\item S. Weinberg, {\it Physica} {\bf 96A} (1979), 327.  For
reviews of more recent work,  see H. Leutwyler, in {\em
Proceedings of the XXVI International Conference on High
Energy Nuclear Physics, Dallas, 1992}, ed. by J. Sanford
(American Institute of Physics, New York, 1193): 185; U. G.
Meissner, {\it Rep. Prog. Phys.} {\bf 56} (1993), 903; A.
Pich, Valencia preprint FTUV/95-4, February 1995, to be
published in {\it Reports on Progress in Physics}; J.
Bijnens, G. Ecker, and  J.  Gasser, in {\it The Daphne
Physics Handbook}, Vol. 1, eds. L. Maiani, G. Pancheri, and
N. Paver (INFN, Frascati, 1995): Chapters 3 and 3.1;~G.
Ecker, preprint hep-ph/9501357, to be published in {\it
Progress in Particle and Nuclear Physics}, Vol. 35 (Pergamon
Press, Oxford).

\item A. Polyakov, {\em Nuc. Phys.} {\bf B268} (1986), 406;
J. Polchinski and A. Strominger, {\em Phys. Rev. Lett.} {\bf
67}, 1681 (1991).

\item C. Becchi, A. Rouet, and R. Stora, {\it Comm. Math.
Phys.} {\bf 42} (1975), 127; in {\em Renormalization
Theory}, ed. by G. Velo and A. S. Wightman (Reidel,
Dordrecht, 1976); {\it Ann. Phys.} {\bf 98} (1976), 287; I.
V. Tyutin, Lebedev Institute preprint N39 (1975).

\item B. W. Lee and J. Zinn-Justin, {\it Phys. Rev.} {\bf
D5} (1972), 3121, 3137; {\it Phys. Rev.} {\bf D7} (1972),
1049; G. `t Hooft and M. Veltman, {\it Nucl. Phys.} (1972)
{\bf B50}, 318; B. W. Lee, {\it Phys. Rev.} {\bf D9} (1974),
933.

\item See, e.g., R. Utiyama and B. De Witt, {\em J. Math.
Phys.} {\bf 3} (1962), 608; S. Weinberg, in {\em Proceedings
of the XVII International Conference on High Energy Nuclear
Physics} (Rutherford Laboratory, 1974), p. III-59;  S.
Deser, in {\em Gauge Theories and Modern Field Theory}, ed.
by R. Arnowitt and P. Nath (M.I.T. Press, Cambridge, 1976);
K. S. Stelle, {\em Phys. Rev.}, {\bf D 16} (1977), 953.

\item J. Zinn-Justin, in {\it Trends in Elementary Particle
Theory - International Summer Institute on Theoretical
Physics in Bonn 1974} (Springer-Verlag, Berlin, 1975).

\item I. A.
Batalin and G. A. Vilkovisky, {\em Phys. Lett.} {\bf B102}
(1981), 27; {\it Nucl. Phys.} {\bf B234} (1984), 106; {\em
J. Math. Phys.} {\bf 26} (1985), 172.    For a review, see
J. Gomis, J. Par\'{\i}s
and S. Samuel, {\it Phys. Rep.} {\bf 259} (1995), 1.

\item G. Barnich and M. Henneaux, {\it Phys. Rev. Lett.}
{\bf 72} (1994) 1588; G. Barnich, F. Brandt, and M.
Henneaux, {\em Phys. Rev.} {\bf 51}, R1435 (1995);
Brussels--Amsterdam preprint ULB-TH-94/07,
NIKHEF-H 94-15, to be published in {\it Comm. Math. Phys};
Brussels--Leuven
preprint KUL-TF-95/16, ULB-TH-95/07.

\item F. Brandt, W. Troost, and A. Van  Proeyen, Leuven
preprint KUL-TF-95/17 (September 1995).

\item G. Bandelloni,  C. Becchi, A. Blasi,  and R. Collina,
{\em Ann. Inst. Henri Poincar\'e A} {\bf 28} (1978), 15.

\item J. Dixon, Nucl. Phys. {\bf B99} (1975), 420 .

\item B. L.
Voronov and I. V. Tyutin, {\em Theor. Math. Phys.} {\bf 50}
(1982), 218; {\bf 52} (1982), 628; B. L. Voronov, P. M.
Lavrov,  and I. V.
Tyutin, {\em Sov. J. Nucl. Phys.} {\bf 36} (1982), 292; P.
M. Lavrov and I. V. Tyutin {\it Sov. J. Nucl. Phys.} {\bf
41} (1985), 1049.

\item D. Anselmi, {\em Class. and Quant. Grav.} {\bf 11}
(1994),
2181; {\bf 12} (1995), 319.

\item M. Harada, T. Kugo, and K. Yamawaki, {\em Prog. Theor.
Phys.} {\bf 91} (1994), 801.

\item M. Henneaux and C. Teitelboim, {\em Quantization of
Gauge Systems} (Princeton University Press, Princeton,
1992): Section 18.1.4; M. Lavrov and I. V. Tyutin, ref. 13.

\item J. Gomis and J. Par\'{\i}s, {\em Nucl. Phys.} {\bf
B341} (1994), 378.

\item J. Gomis, K. Kamimura, and R. Kuriki Barcelona-Tokyo
preprint UB-ECM-PF 95/22, TOHO-FP-9553, to be published
(1995).

\end{enumerate}
\end{document}